\documentclass[sn-mathphys-ay]{sn-jnl}% Math and Physical Sciences Author Year Reference Style
\usepackage{graphicx}%
\usepackage{multirow}%
\usepackage{amsmath,amssymb,amsfonts}%
\usepackage{amsthm}%
\usepackage{mathrsfs}%
\usepackage[title]{appendix}%
\usepackage{xcolor}%
\usepackage{textcomp}%
\usepackage{manyfoot}%
\usepackage{booktabs}%
\usepackage{algorithm}%
\usepackage{algorithmicx}%
\usepackage{algpseudocode}%
\usepackage{listings}%
\usepackage{bbm}
\usepackage{float}
\usepackage{mathtools}
\usepackage{indentfirst}
\usepackage{makecell}
\usepackage{diagbox}
\usepackage{dsfont}
\usepackage{color,soul}
\usepackage{hyperref}
\hypersetup{
    colorlinks=true,
    linkcolor=blue,     
    }
\begin{document}

\title[Expectile regression averaging]{Expectile regression averaging method for probabilistic forecasting of electricity prices}


\author*[1]{Joanna Janczura %\orcidlink{0000-0001-6110-3394}
}\email{joanna.janczura@pwr.edu.pl}

\affil*[1]{\orgdiv{Faculty of Pure and Applied Mathematics, Hugo Steinhaus Center}, \orgname{Wroc{\l}aw University of Science and Technology}, %\orgaddress{\street{Wyb. Wyspia\'nskiego 27}, \city{Wroc{\l}aw}, \postcode{50-370}, 
\country{Poland}}
%}


\abstract{In this paper we propose a new method for probabilistic forecasting of electricity prices. It is based on averaging point forecasts from different models combined with expectile regression. We show that deriving the predicted distribution in terms of expectiles, might be in some cases advantageous to the commonly used quantiles. We apply the proposed method to the day-ahead electricity prices from the German market and compare its accuracy with the Quantile Regression Averaging method and quantile- as well as expectile-based historical simulation. The obtained results indicate that using the expectile regression improves the accuracy of the probabilistic forecasts of electricity prices, but a variance stabilizing transformation should be applied prior to modelling.  }

\keywords{electricity market, expectile regression, probabilistic forecast, forecast averaging}

\maketitle

\section{Introduction}
For the last three decades electricity markets are undergoing significant structural changes. 
 At the same time the price risk for the wholesale electricity market participants increased significantly. Limited storage possibilities, technical constraints of transmission grid and the importance of electricity supply lead to much higher price variability than in other commodities markets. In the recent years we observe a rapid transformation of the overall electricity production profile in the European electricity markets with a growing share of renewable energy sources (RES). This makes not only electricity demand but also supply highly weather dependent and, as a consequence, electricity prices can be even more volatile. 
The distributed generation caused by the growth of RES induced the change not only in the energy production, but also in the profiles of the market participants. A number of small producers and traders joined the market. They are facing a significant risk associated with electricity price volatility, but at the same time electricity markets give them a range of trading opportunities. Since electricity prices are not known in advance, any trade planing needs to be based on price forecasts. In such a context some trading strategies have been proposed recently in the literature \citep{maciejowska,SERAFIN_ID3, Petukhina}. They utilize point as well as probabilistic forecasts of electricity prices. We believe that improving the accuracy of the forecasts would also improve the efficiency of trading strategies.  

There are many articles dedicated to point forecasting of the day-ahead electricity prices, see \cite{RWreview} or \cite{MarcjaszWeron21} for extensive reviews. %Recently,  
Also an extension from the point to probabilistic forecasting methods has gained much attention in recent years, see \cite{nowotarski_weron_Review} for a review. The latter takes into account not only the best estimate of a future value but also uncertainty of the prediction. As a consequence, it brings much more information to a decision maker and allows e.g. for a direct risk management. One of the methods that were successfully applied in the probabilistic electricity price forecasting is the Quantile Regression Averaging (QRA) proposed by \cite{nowotarski}. It is built on the quantile regression \citep{RK} method combined with different point forecasts of electricity prices. In this paper we follow this direction and introduce the Expectile Regression Averaging (ERA) method. It uses a notion of expectiles introduced originally by \cite{newey}. Expectiles can be viewed as an analogous description of the distribution to quantiles \citep{gneiting}. However, the estimation of the expectile regression is based on the least squares method, in contrast to the quantile regression which is based on the least absolute deviations. 

Due to their numerical and statistical properties expectiles has seen increasing interest in the recent years. They were used, among others, in regression analysis \cite{expectile:waltrup}, functional factor modelling \citep{expectile_func}, estimation of extremes \citep{expectile_extreme} or multivariate data analysis \citep{expectile_bivar}. Expectiles have also gained at lot of attention in finance, since \cite{kuan} adopted them as a risk measure, called the expectile Value at Risk (EVaR). Although the interpretation of EVaR is less straightforward than for the classical risk measures, using expectiles allows to overcome the known drawbacks of the latter, like non-coherence or non-elicitability \citep{Ziegel, Bellini}. Expectiles were also recently used as a risk measure for electricity market by \cite{expectiles_electricity} or \cite{JANCZURAWOJCIK2022}. For other applications concerning electricity markets see also the work of \cite{TAYLOR2021800} or \cite{Hardle_expectiles}. However, to our best knowledge, expectiles were not used in the context of forecast averaging for electricity prices, yet.

The rest of the paper is structured as follows. In Section \ref{sec:expectiles} we briefly describe the notion of expectiles and show their analogies as well as differences from quantiles. Section \ref{sec:forecasts} is devoted to the construction of probabilistic forecasts of electricity prices. In particular, in Section \ref{sec:ERA} we introduce the Expectile Regression Averaging method. Next, in Section \ref{sec:results} we apply the proposed technique to the day-ahead electricity prices from the German market and compare its performance with some benchmark probabilistic forecasts. % based on the historical simulation method, 
Finally, in Section \ref{sec:conclusions} we conclude. 

\section{Expectiles and quantiles} \label{sec:expectiles}
A standard way of describing the probability distribution of a random variable is in terms of the cumulative distribution function (CDF) and its inverse, i.e. quantiles. Another notion that can be used in such a context is the expectile. An expectile at level $\tau$, $e_{\tau}$ ($0<\tau<1$), is defined as a unique solution of \citep{newey}
\begin{equation}
\label{eq:expectile_def}
 \tau\mathbb{E}[(Y-e_{\tau}(Y))_{+}] = (1-\tau)\mathbb{E}[(Y-e_{\tau}(Y))_{-}],
\end{equation}
where  $(x)_+=\max(x,0)$ and $(x)_-=\min(x,0)$ denote the positive and negative part of a variable x. For $\tau=\frac{1}{2}$ expectile is equal to the mean of the distribution, so expectiles are often seen as asymmetric generalization of the mean  \citep{gneiting}. On the other hand, if the expected value in (\ref{eq:expectile_def}) is replaced with a probability mass function, then the formula defines the quantile, yielding median for $\tau=\frac{1}{2}$. Hence, expectiles generalize the mean in a similar way as quantiles generalize the median, but are based on the mean distance instead of the mass of the distribution. As a consequence, they include information on the size of exceedances, in contrast to quantiles, which are based only on their frequency. 
%As a consequence a given expectile is linked to the whole distribution and 
%$    \tau=\frac{\mathbb{E}[|Y-e_{\tau}(Y)| \mathbbm{1}_{Y<e_{\tau}(Y)}] }{\mathbb{E}[|Y-e_{\tau}(Y)|] },
%$
%while the quantile at level $\alpha$, $q_{\alpha}$, only to its tail
%$
%    \alpha=\text{P}(Y<q_{\alpha} )=\mathbb{E}[ \mathbbm{1}_{Y<q_{\alpha}(Y)}].
%$

Expectiles can be also defined as the minimizers of the quadratic loss function \citep{Bellini}     
\begin{equation}
\label{eq:expectile}
e_{\tau}(Y)= \arg\min_{x\in \mathbb{R}} \tau \mathbb{E}[(Y -x)^2_{+}] + (1-\tau)\mathbb{E}[(Y -x)^2_{-}]. 
\end{equation}
Note that for $\tau=\frac{1}{2}$ this loss function is just the standard mean square error (MSE). For quantiles we have an analogous absolute loss function
\begin{equation}
\label{eq:quantile}
q_{\alpha}(Y)= \arg\min_{x\in \mathbb{R}} \alpha \mathbb{E}[|Y -x|_{+}] + (1-\alpha)\mathbb{E}[|Y -x|_{-}],
\end{equation}
which for $\alpha=\frac{1}{2}$ is the mean absolute error (MAE). The loss  functions (\ref{eq:expectile}) and (\ref{eq:quantile}) are also a basis for the quantile and expectile regression, generalizing the classical linear regression model in terms of the predicted variable distribution. These methods will be further used in the paper for forecast construction.  

Both quantiles and expectiles describe a distribution of a random variable. Naturally, they are related with each other. As shown by \cite{expectile:yao} there exists a unique function $h$ such that 
\begin{equation}
    q_{\alpha}(Y)=e_{h(\alpha)}(Y).  
\end{equation}
It is given by 
\begin{equation}
\label{eq:h_function}
    h(\alpha)=\frac{-\alpha q_{\alpha}(Y)+G(q_{\alpha}(Y))}{-e_{0.5}(Y)+2G(q_{\alpha}(Y))+(1-2\alpha)q_{\alpha}(Y)},
\end{equation}
were $G(x)=\int_{-\infty}^x ydF(y)$ is the partial moment function and $F$ is the CDF of $Y$. There exists also the inverse relation. Expectiles are linked with the CDF $F$ by \citep{expectile:waltrup}
\begin{equation}
   \label{eq:expectile_cdf}
   e_{\tau}(Y)=\frac{(1-\tau)G(e_{\tau}(Y))+\tau (e_{0.5}(Y)-G(e_{\tau}(Y)))}{(1-\tau)F(e_{\tau}(Y))+\tau (1-F(e_{\tau}(Y)))}.
\end{equation}
Hence, quantiles can be calculated from expectiles, and expectiles can be calculated form quantiles, but it usually requires some numerical approximations.

\section{Probabilistic forecasts of electricity prices}\label{sec:forecasts}
\subsection{Quantile and Expectile regression averaging} \label{sec:ERA}
%\subsection{Forecast calculation}\label{sec:ERA}
One of the commonly used methods for probabilistic orecasting of electricity prices is the Quantile Regression Averaging (QRA) proposed by \cite{nowotarski}. It is based on applying the  quantile regression \citep{RK} to a pool of point forecasts of different individual models. Denote electricity price for a delivery during hour $h$ on day $t$ by $P_{h,t}$. 
In the QRA method, probabilistic forecasts of $P_{h,t}$ are determined as the following linear combination \citep{nowotarski}
\begin{equation}
\label{eq:quant_reg1}
\hat{q}_{P_{h,t}}\left(\alpha\right) =\hat{\mathbf{P}}_{h,t}\mathbf{w}_{\alpha},
\end{equation}
where $\hat{q}_{P_{h,t}}(\alpha)$ is an $\alpha$-quantile of the forecasted distribution, $\hat{\mathbf{P}}_{h,t}$ is a vector of $K$ corresponding individual point forecasts, while $\mathbf{w}_{\alpha}$ is a column of weights for the $\alpha$-quantile. Weights $\mathbf{w}_{\alpha}$ are estimated, by minimizing the quantile loss function 
\begin{equation}
\label{eq:quant_reg2}
\min_{\mathbf{w}_{\alpha}} \left[ \sum_{t=1}^T\left( \alpha\left|P_{h,t} -  \hat{\mathbf{P}}_{h,t}\mathbf{w}_{\alpha}\right| \mathds{1}_{\{P_{h,t} \geq \hat{\mathbf{P}}_{h,t}\mathbf{w}_{\alpha}\}}+ (1-\alpha) \left|P_{h,t} -\hat{\mathbf{P}}_{h,t}\mathbf{w}_{\alpha}\right|\mathds{1}_{\{P_{h,t} < \hat{\mathbf{P}}_{h,t}\mathbf{w}_{\alpha}\}}\right) \right].
\end{equation}

In this paper we follow the forecast averaging approach, but we propose to combine it with the expectile regression. It is similar to the quantile regression, but the absolute loss function (\ref{eq:quant_reg2}) is replaced with the quadratic one (\ref{eq:expectile}). Hence, in the Expectile Regression Averaging (ERA) method the $\tau$-expectile of the predicted distribution, $\hat{e}_{P_{h,t}}\left(\tau\right)$,  is calculated as  
\begin{equation}
\label{eq:exp_reg1}
\hat{e}_{P_{h,t}}\left(\tau\right) =\hat{\mathbf{P}}_{h,t}\mathbf{w}_{\tau},
\end{equation}
where $\hat{\mathbf{P}}_{h,t}$ is the vector of point forecasts from the individual models and $\mathbf{w}_{\tau}$ are the weights estimated from
\begin{equation}
\label{eq:exp_reg2}
\min_{\mathbf{w}_{\tau}} \left\{ \sum_{t=1}^T\left[ \tau\left(P_{h,t} -  \hat{\mathbf{P}}_{h,t}\mathbf{w}_{\tau}\right)^2 \mathds{1}_{\{P_{h,t} \geq\hat{\mathbf{P}}_{h,t}\mathbf{w}_{\tau}\}}+ (1-\tau) \left(P_{h,t} - \hat{\mathbf{P}}_{h,t}\right)^2\mathds{1}_{\{P_{h,t} < \hat{\mathbf{P}}_{h,t}\mathbf{w}_{\tau}\}}\right] \right\}.
\end{equation}
 Note that the expectile regression is based on the $L_2$ optimization, yielding here an ordinary least squares (OLS) method, while the quantile regresssion is based on $L_1$ optimization. 
 The latter is more robust to outliers, but on the other hand least squares method posses better numerical properties.

\subsection{Individual models}
The ERA and QRA methods use a linear combination of individual forecasts, so they require deriving a set of point forecasts, first. To this end, we consider five expert models, being standard, frequently used approaches in electricity price modelling \citep[see e.g.][]{Misiorek2006, Kristiansen, Maciejowska_EE}. All are build on autoregressive models with exogenous variables (ARX), in which one assumes that electricity prices can be explained by the market fundamentals of technical or economical nature, like e.g. load, generation or weather conditions. Since, the forecasts of physical system variables are often publicly available, the construction of price predictions with the ARX models is straightforward. 

In the first considered in this paper model we assume that the electricity price for a delivery during hour $h$ of day $t$, $P_{h,t}$, is given by  
\begin{equation}
\label{M1}
\text{Model 1:}\quad\quad  P_{h,t} =  \theta_{1} P_{h,t-1}+\theta_{2} P_{h,t-2}+\theta_{7} P_{h,t-7} + \sum_{i=1}^k \psi_i Z_{h,t}^i + \sum_{i=1}^4\alpha_i D_t + \epsilon_{h,t},
\end{equation}
where $P_{h,t-i}$ are the autoregressive terms, $Z_{h,t}^i, i=1,2,..,k$ are the exogenous variables and $\epsilon_{h,t}$ is the noise term. In order to account for a weekly seasonality of electricity prices, we use also a dummy variable $D_t$ related to different days of the week. Here we consider Monday, Saturday, Sunday/Holiday, and the other days of the week. 

The second model differs from the first one by the number of regressors, for which we consider all prices from a given hour during the previous week, i.e. 
\begin{equation}
\label{M2}
\text{Model 2:}\quad\quad   P_{h,t} =  \sum_{i=1}^7\theta_{i} P_{h,t-i} + \sum_{i=1}^k \psi_i Z_{h,t}^i + \sum_{i=1}^4\alpha_i D_t + \epsilon_{h,t}.
\end{equation}

The third model uses also the minimum and maximum of the previous days' prices, so it allows for taking into account nonlinear intraday effects\vspace{10pt}
\\

Model 3: 
\begin{equation}
\label{M3}
P_{h,t} =  \sum_{i=1}^7\theta_{i} P_{h,t-i} + \sum_{i=1}^k \psi_i Z_{h,t}^i + \sum_{i=1}^4\alpha_i D_t + \delta \min_h(P_{h,t-1}) + \eta \max_h(P_{h,t-1}) +  \epsilon_{h,t}.
\end{equation}

The structure of the fourth model, called the p-ARX \citep{Misiorek2006}, is similar to Model 1, but applied to prices with pre-processed spikes. 
Precisely, the prices that exceed the mean level from the calibration window by more than its three standard deviations are transformed as
\begin{align}
P^p_{h,t} =  
\begin{cases}
L_U+L_U\log_{10}\left(\left| \frac{P_{h,t}}{L_U}\right| \right)  & \text{if} \quad P_{h,t}>L_U,\\ 
L_L-\lvert L_L\rvert \log_{10}\left(\left| \frac{P_{h,t}}{L_L}\right| \right)  & \text{if} \quad P_{h,t}<L_D, 
\end{cases}
\end{align}
where the upper level is set to $L_U= \mu_{P_{h,t}} + 3 \sigma_{P_{h,t}}$, while the lower level to $L_L= \mu_{P_{h,t}} - 3 \sigma_{P_{h,t}}$.

The fifth model specification, m-ARX proposed by \cite{ZIEL2018}, is a modification of Model 2, including the weekly mean of the prices $\overline{P}^W_{h,t}=\frac{1}{7}\sum_{i=1}^7 P_{h,t-i} $ in the following way  
\begin{equation}
\label{M5}
\text{Model 5:}\quad\quad  P_{h,t} = \overline{P}^W_{h,t} +  \sum_{i=1}^7\theta_{i}( P_{h,t-i} - \overline{P}^W_{h,t}) + \sum_{i=1}^k \psi_i Z_{h,t}^i + \sum_{i=1}^4\alpha_i D_t + \epsilon_{h,t}.
\end{equation}

The parameters of the ARX models, $\theta_i, \alpha_i, \psi_i,\delta,\eta$, can be estimated using the least squares method. Then, the day-ahead point forecasts for each hour are given by the corresponding linear combination of explanatory variables. The set of these forecasts, $\mathbf{\hat{P}}_{h,t}$, is then used in the ERA (\ref{eq:exp_reg1}) and QRA (\ref{eq:quant_reg1}) methods.  

As a benchmark we also calculate the probabilistic forecasts using the standard historical simulation method. For each of the individual models we derive the out-of-sample point prediction errors  
\begin{equation}
\epsilon_{h,t}= \hat{P}_{h,t} - P_{h,t}.
\end{equation}
Then, the probabilistic forecast is calculated as the sum of the point forecast and the errors' distribution. Here, this distribution is considered in terms of the quantiles as well as expectiles. 

Overall, we consider 12 methods for deriving probabilistic forecasts: QRA, ERA as well as historical simulation of expectiles and quantiles from the five individual models. The models are fitted for each hour separately, so in total we consider 24 one-dimensional time series. This is a common approach in electricity price modelling since electricity delivered during different hours is in fact traded as separate products.  

\subsection{Prediction intervals from expectiles}
The considered probabilistic forecasts are given either in terms of quantiles or of expectiles. Both are a proper description of the predicted distribution, but their accuracy should be evaluated using different scoring functions. Hence, in order to compare the quantile- and expectile-based methods, we transform expectiles into the corresponding quantiles. This yields the  prediction intervals (PI), commonly used in forecasting context. To this end, we use a procedure proposed by \cite{expectile:waltrup}. It is based on finding a CDF that minimizes the distance between the derived expectiles and their theoretical values resulting from that CDF (\ref{eq:expectile_cdf}), i.e.
\begin{equation}
   \label{eq:expectile_cdf_num}
   \arg\min_F \left[\hat{e}_{\tau}(Y)-\frac{(1-\tau)G(e_{\tau}(Y))+\tau (e_{0.5}(Y)-G(e_{\tau}(Y)))}{(1-\tau)F(e_{\tau}(Y))+\tau (1-F(e_{\tau}(Y)))}\right]^2, 
\end{equation}
where $F(e_{\tau}(Y))$ is the value of the CDF at $\tau$-expectile and $G(e_{\tau}(Y))$ is its partial moment function, $G(x)=\int_{-\infty}^x ydF(y)$. Next, the values of the CDF at desired quantiles are approximated using linear interpolation and finally inverted yielding prediction intervals. For a detailed description of this procedure see \cite{expectile:waltrup}.

\subsection{Variance stabilizing transformation}
Since electricity prices are known to be highly volatile, we apply the variance stabilizing transformation prior to modelling \citep[see][for a discussion on a usage of different transformations in this context]{uniejewski_vst}. Here, we apply the inverse hyperbolic sine (asinh) function, which can be viewed as a generalization of the logarithmic transformation, being suitable also for negative prices. Namely, we consider 
\begin{equation}
    P_{h,t} = \text{asinh}(y_{h,t}) = \log\left(y_{h,t} + \sqrt{(y_{h,t})^2 + 1}\right), 
    \label{eq:asinh}
\end{equation}
where $y_{h,t}$ is the normalized price, $y_{h,t} = {\left ( p_{h,t} - \mu_{p_{h,t}} \right)/\sigma_{p_{h,t}}}$,    
with $\sigma_{p_{h,t}}$ being here the standard deviation of prices, $p_{h,t}$, in the calibration window and $\mu_{p_{h,t}}$ the corresponding mean.

For practical applications one is usually interested in predictions of the original prices, hence, here, predictions calculated for transformed prices are in the end inverted back. Since inverting the asinh transformation of random variables is not straightforward (see \cite{Narajewski} for a discussion on this issue), we use the Monte Carlo approach. Namely, first we simulate $n$ day-ahead price scenarios using the predicted distribution. Next, we invert each of them using the hyperbolic sine function 
\begin{equation}
    {p}^j_{h,t}= \sigma_{p_{h,t}} \cdot \text{sinh}\left( \hat{P}^j_{h,t}  \right) + \mu_{p_{h,t}} , j=1,2,...,n. 
    \label{eq:inverse_asinh}
\end{equation}
Finally, the empirical distribution of the inverted scenarios $p^1_{h,t},p^2_{h,t},...,p^n_{h,t}$ yields the probabilistic forecast of the price for day $t$ and hour $h$. 

\section{German electricity market case study}\label{sec:results}
\subsection{Datasets}
\begin{figure}[h]
    \centering
    \includegraphics[scale=0.75]{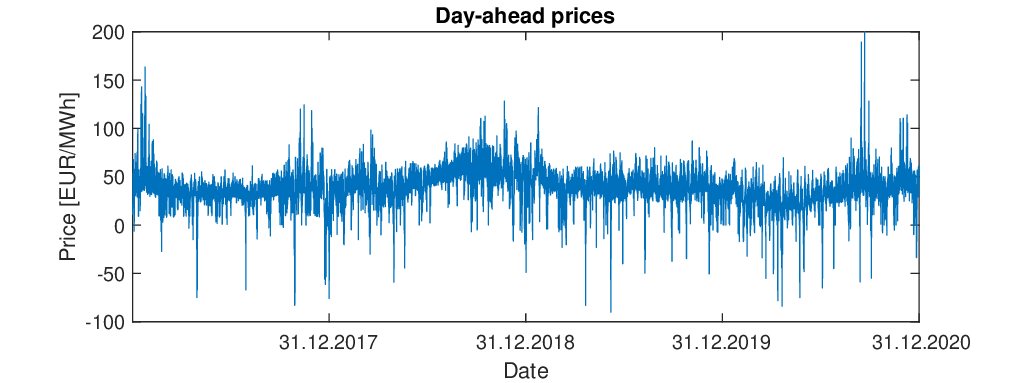}
    \caption{Hourly day-ahead electricity prices from the German EPEX spot market from the period 1.01.2017-31.12.2020}
    \label{fig:data}
\end{figure}

\begin{figure}[h]
    \centering
    \includegraphics[scale=0.75]{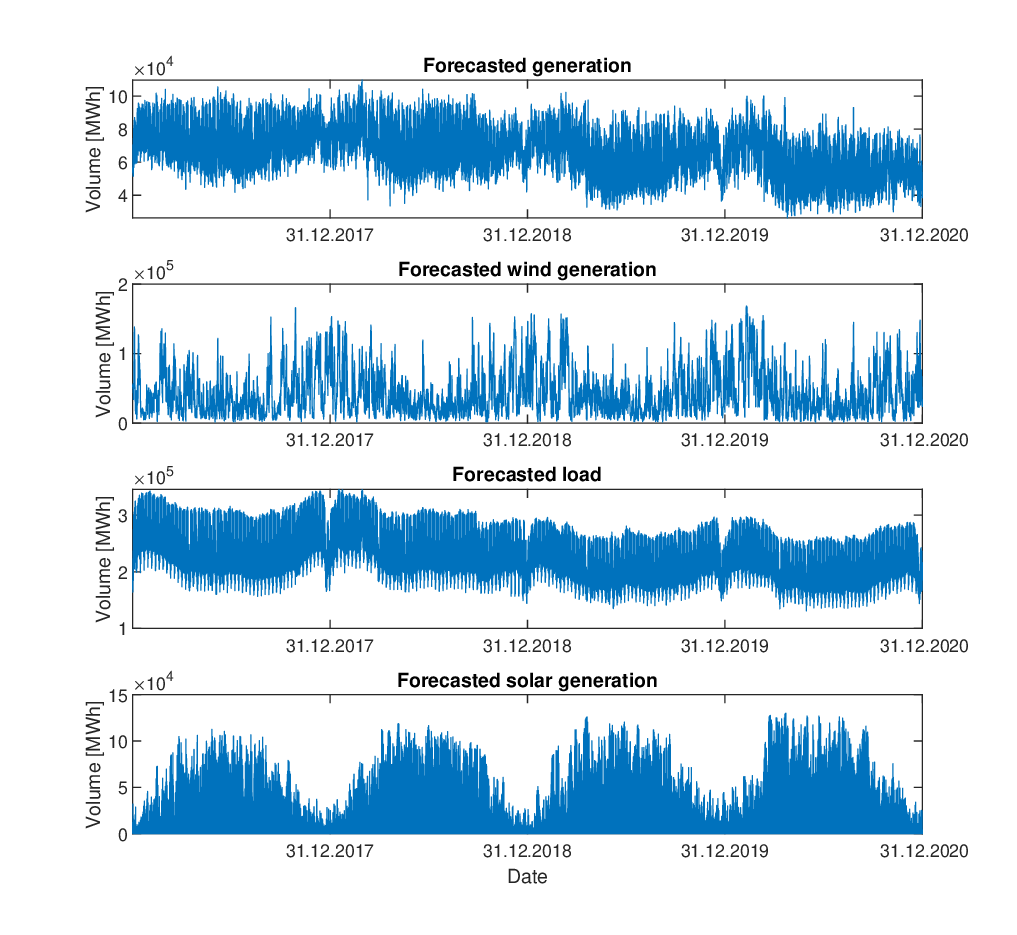}
    \caption{Values of the exogenous variables: forecasted generation, forecasted wind generation, forecasted load and forecasted solar generation for the German market from the period 1.01.2017-31.12.2020 (source: ENTSO-E).}
    \label{fig:ex_data}
\end{figure}

We apply the ERA, QRA as well as the historical simulation methods from individual models (\ref{M1})-(\ref{M5}) to the day-ahead hourly electricity prices from the German EPEX spot market spanning the period of 1.01.2017-31.12.2020. The considered prices are plotted in Figure \ref{fig:data}. For the calculation of the point forecasts we use the set of exogenous variables $Z^i_{h,t}$ consisting of: i) the forecasts of generation; ii) forecasts of wind generation; iii) forecasts of solar generation and iv) forecasts of load. All these values are published by the Transmission System Operator (TSO) and are freely available from ENTSO-E platform (https://transparency.entsoe.eu/). The values of the considered variables are plotted in Figure \ref{fig:ex_data}. 

\subsection{Forecasts construction}
Electricity price predictions are calculated in a moving window scheme. For each day of the validation window we calculate the day-ahead probabilistic forecast based on the parameters estimated from the preceding calibration window. The derivation of the probabilistic forecasts for all considered methods requires calculating the point forecasts, first. Hence, we divide the calibration window into two yearly parts. The first one is used for the estimation of the parameters of the individual models (\ref{M1})-(\ref{M5}). Next, the resulting point forecasts are derived for the second part of the calibration window. Finally, these point forecasts are used to calculate probabilistic forecasts for the validation window. Here, the forecasts are evaluated in a two-yearly window spanning over the years 2019-2020. The comparison of forecasts is done in terms of quantiles (prediction intervals). In order to transform the expectile-based predictions, we apply the procedure (\ref{eq:expectile_cdf}) to a grid of expectiles calculated at the following levels $\tau = 0.001,0.0025,0.005,0.0075,0.01,0.02,0.04,...,0.98,0.99,$ $0.9925,0.995,0.9975,0.999$. 

\subsection{Forecasts evaluation}
\begin{figure}[h]
    \centering
    \includegraphics[scale=0.75]{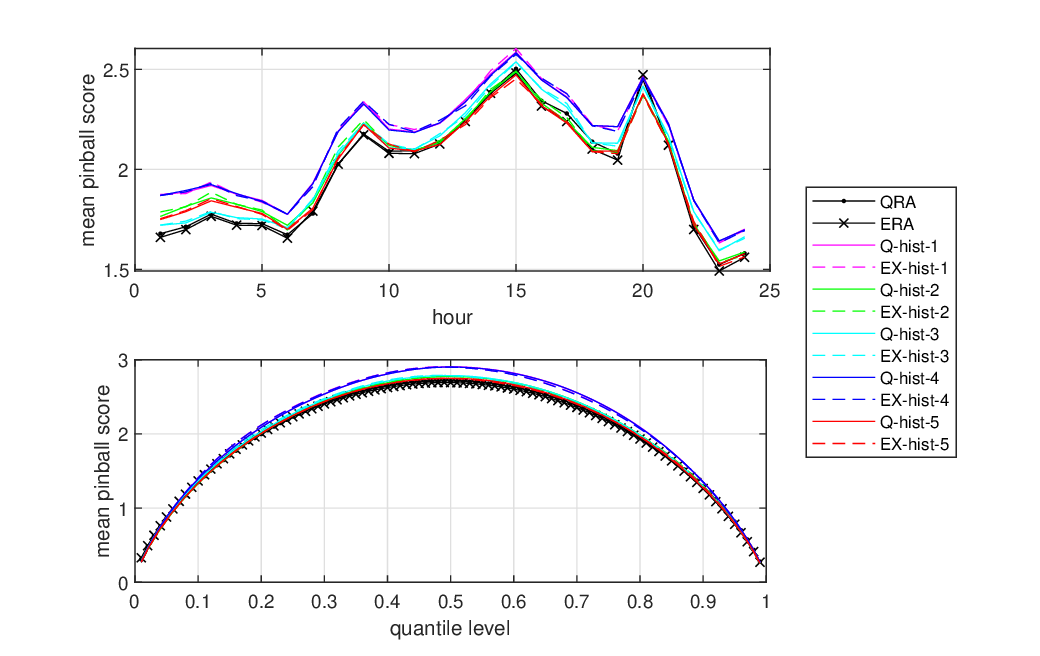}
    \caption{Hourly (top panel) or percentile (bottom panel) mean pinball score obtained for each of the considered models applied to asinh transformed prices. 'Q-hist' denote the historical simulation in terms of quantiles, while 'EX-hist' the historical simulation in terms of expectiles. Numbers are related to the individual Models.} 
    \label{fig:PL}
\end{figure}

The accuracy of prediction intervals is compared using the pinball loss (PL), being a proper scoring function for quantiles, \citep{Gneiting_PF}
\begin{equation}
PL\left(\hat{q}_{P_{t,h}}(\alpha),P_{t,h},\alpha\right)=\left\{\begin{array}{lcr}
(1-\alpha)\left(\hat{q}_{P_{t,h}}(\alpha) - P_{t,h}\right) & \mbox{if} & P_{t,h}<\hat{q}_{P_{t,h}}(\alpha),\\
\alpha\left( P_{t,h} - \hat{q}_{P_{t,h}}(\alpha)\right) & \mbox{if} & P_{t,h}\geq\hat{q}_{P_{t,h}}(\alpha),\\
\end{array}\right.
\end{equation} 
 where $\hat{q}_{P_{t,h}}(\alpha)$  is the $\alpha$-quantile of the forecasted price distribution and $P_{t,h}$ is the actually observed value. We calculate the averaged pinball score for each hour and percentile in the validation window. The values are then averaged over all percentiles, yielding hourly pinball score, or over all hours, yielding percentile pinball score. The obtained results are plotted in Figure \ref{fig:PL}. As can be observed, the ERA ad QRA averaging schemes yield lower pinball scores than the historical simulation method for most percentiles. The highest difference is obtained in the middle of the distribution. A similar picture is obtained for the hourly pinball score with the lowest values for the ERA method for most of the hours. The only exceptions are the peak hours of 20 and 15, for which the historical simulation methods yield more accurate results than the forecast averaging.  

\begin{figure}[h]
    \centering
    \includegraphics[scale=0.75]{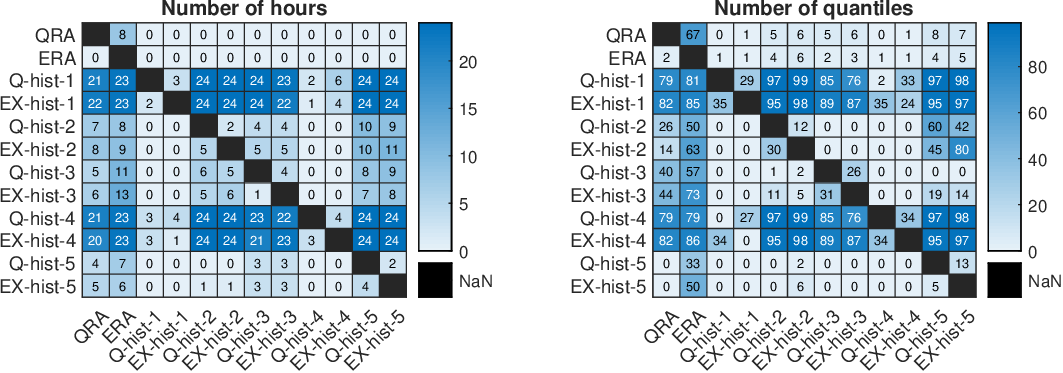}
    \caption{Number of hours (left panel) or percentiles (right panel) for which prediction from the model in row is significantly worse than prediction from the model in column according to the \cite{DM_test} test.  'Q-hist' denote the historical simulation in terms of quantiles, while 'EX-hist' the historical simulation in terms of expectiles. Numbers are related to the individual Models. The test was performed at the $5\%$ significance level.}
        %\vspace{-3.7cm}
    \label{fig:DM}
\end{figure}

The significance of the pinball score differences is further verified using the one-sided \cite{DM_test} test. In Figure \ref{fig:DM} we show the number of hours as well as percentiles for which each of the considered models was significantly outperformed by its competitors. The obtained results confirm conclusions drawn from Figure \ref{fig:PL}. The ERA and QRA averaging schemes outperform significantly the historical simulation methods. There are no hours for which the accuracy of ERA or QRA were significantly lower, while they outperformed the historical simulation for 5 up to 23 hours depending on the model specification. Similarly for percentiles, we can see a significant improvement in the forecast accuracy, if averaging methods were used. This is especially apparent for the ERA method, which outperforms the other approaches for 33 up to 86 percentiles. It yields significantly better results also in comparison with the QRA method, outperforming the latter for 8 hours and 67 percentiles. Looking at the differences between the quantile- and expectile-based historical simulation within a given model specification we do not observe a clear pattern and overall accuracy is at similar level. 
 
\begin{figure}[h]
    \centering
    \includegraphics[scale=0.75]{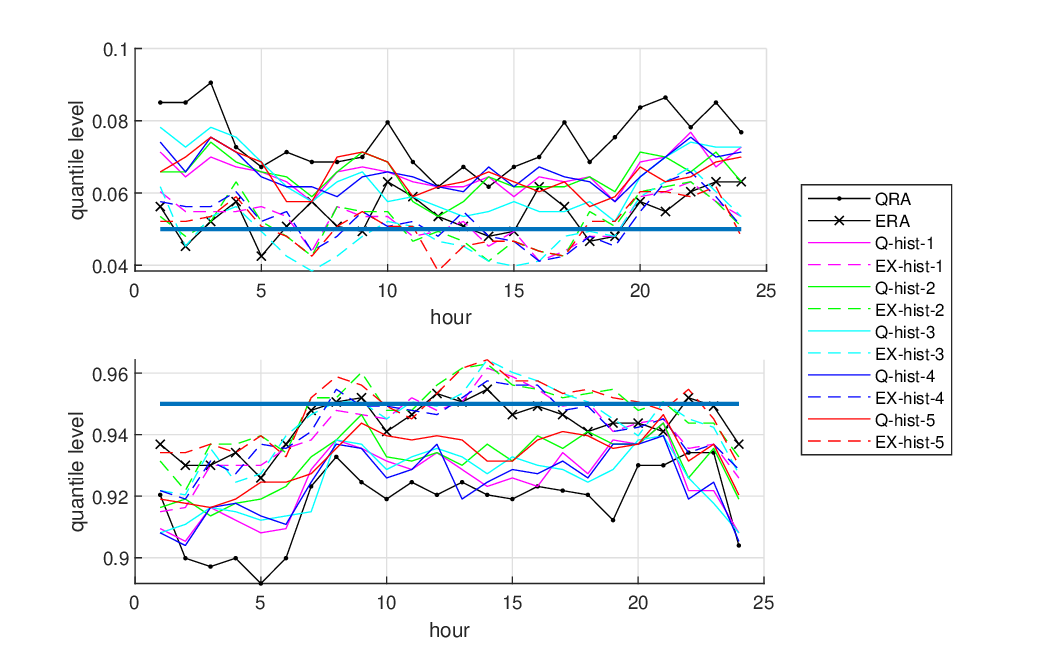}
    \caption{Hourly coverage probabilities at the $5\%$ (top panel) and $95\%$ (bottom panel) level obtained for each of the considered models applied to  asinh transformed prices. 'Q-hist' denote the historical simulation in terms of quantiles, while 'EX-hist' the historical simulation in terms of expectiles. Numbers are related to the individual Models. The $5\%$ and $95\%$ levels are marked with horizontal blue lines.}
    \label{fig:cov}
\end{figure}

 In order to further evaluate the predictions, we calculate the coverage probability $\text{P}(P_{t,h}< \hat{q}_{P_{t,h}}(\alpha))$ at the $5\%$ and $95\%$ $\alpha$-levels. Note that these are in fact accuracy of the Value at Risk forecasts at $95\%$ level, i.e. VaR$_{95\%}$, for a seller and buyer, respectively. Results obtained for each of the hours are plotted in Figure \ref{fig:cov}. Here, we can see a clear difference between the results obtained with the quantile- and expectile-based methods. The coverage probabilities obtained for the latter are closer to the expected $5\%$ and $95\%$ levels. This is visible for both approaches - the ERA method and the expectile-based historical simulation. The coverage probabilities obtained with the QRA as well the quantile-based historical simulation methods are too high for the $5\%$ quantile and at the same time too low for the $95\%$ one, yielding too narrow prediction intervals. The coverage probabilities obtained with the expectile based methods are close to the expected $5\%$ and $95\%$ levels with the only exception for higher quantiles in the night hours which are lower by approximately $1\%$. 
 The significance of the differences from the expected $5\%$ and $95\%$ levels is verified using the \cite{kupiec} test. The number of hours for which the obtained coverage probabilities were significantly different from $5\%$ and $95\%$ is given in Table \ref{tab:cov_h}. In the case of the expectile-based methods the obtained values are significantly different than the expected ones only for few hours, mainly during night, and for $95\%$ level. For the quantile-based methods the accuracy is much worse as the number of hours with significant differences is from 10 up to even 22.      
\begin{table}[h]
    \centering
    \scriptsize
    \setlength{\tabcolsep}{3pt} 
    \renewcommand{\arraystretch}{1.5}
    \caption{Number of hours for which the coverage probability obtained for the considered methods applied to asinh transformed prices was not significantly different than the expected $5\%$ or $95\%$ value according to the \cite{kupiec} test performed at the $5\%$ significance level. 'Q' denotes the quantile-based approach, while 'EX' the expectile-based one.}
    \begin{tabular}{|c|c|c|c|c|c|c|c|c|c|c|c|c|}
    \hline
    &\multicolumn{2}{|c|}{}&\multicolumn{10}{|c|}{Historical simulation}\\ \hline
    &\multicolumn{2}{|c|}{}&\multicolumn{2}{|c|}{Model 1}&\multicolumn{2}{|c|}{Model 2}&\multicolumn{2}{|c|}{Model 3}&\multicolumn{2}{|c|}{Model 4}&\multicolumn{2}{|c|}{Model 5}\\ \hline
    & QRA & ERA & Q & EX& Q & EX& Q & EX& Q & EX& Q & EX\\
\hline\hline
P$_{5\%} [\%]$&21 & 0 &10 & 0 &13 & 0 &11 & 1 & 11 & 1 & 13 & 0\\
P$_{95\%} [\%]$&22 &3 & 22   &  7  &  14  &   2 &   17  &   6  &  20   &  8  &  14 &    2\\
\hline
    \end{tabular}
    \label{tab:cov_h}
\end{table}

The obtained results are summarized in Table \ref{tab:results}. As a reference we provide also the results obtained in the case, if there was no transformation applied to electricity prices prior to modelling. The coverage probabilities and the pinball scores are averaged over all hours and in the latter case also over all percentiles. The coverage probabilities are additionally evaluated with the \cite{kupiec} test at the $5\%$ significance level. 
\begin{table}[h]
    \centering
    \scriptsize
    \setlength{\tabcolsep}{3pt} 
    \renewcommand{\arraystretch}{1.5}
    \caption{The values of the pinball score (PS) as well as the coverage probabilities P$_{5\%}$ and P$_{95\%}$ at the $5\%$ and $95\%$ levels obtained for the considered methods. The values were averaged over 24 hours and all percentiles. Predictions were calculated for prices transformed with asinh function, as well as without a transformation. 'Q' denotes the quantile-based approach, while 'EX' the expectile-based one. The coverage probabilities that are not significantly different than the expected level according to the \cite{kupiec} test and the lowest pinball scores are given in bold.}
    \begin{tabular}{|c|c|c|c|c|c|c|c|c|c|c|c|c|}
    \hline
    &\multicolumn{2}{|c|}{}&\multicolumn{10}{|c|}{Historical simulation}\\ \hline
    &\multicolumn{2}{|c|}{}&\multicolumn{2}{|c|}{Model 1}&\multicolumn{2}{|c|}{Model 2}&\multicolumn{2}{|c|}{Model 3}&\multicolumn{2}{|c|}{Model 4}&\multicolumn{2}{|c|}{Model 5}\\ \hline
    & QRA & ERA & Q & EX& Q & EX& Q & EX& Q & EX& Q & EX\\
    \hline\hline
     \multicolumn{13}{|c|}{With asinh transformation}\\
\hline\hline
P$_{5\%} [\%]$&7.54&	\textbf{5.26}	&6.64&	\textbf{5.16}	&6.41	&\textbf{5.10}	&6.45&	\textbf{5.01}&	6.55&	\textbf{5.23}	&6.54& \textbf{5.15}\\
P$_{95\%} [\%]$&91.78&	94.28&	92.40	&94.12&	93.04&	\textbf{94.63}&	92.47	&94.36&	92.37&	94.15&	93.23&\textbf{94.83}\\
PS &2.00&	\textbf{1.98}&	2.12& 2.12&	2.02&	2.03&	2.04&	2.04&	2.11&	2.11&	2.01 &2.01\\
\hline\hline
 \multicolumn{13}{|c|}{Without asinh transformation}\\
\hline\hline
P$_{5\%} [\%]$&8.83&	8.02&	7.28	&7.27&	7.19&	7.24&	6.92&	7.10&	6.69&	6.73&	7.21&7.26\\
P$_{95\%} [\%]$&93.61&	94.16&	93.72&	93.67&	94.43	&94.40&	94.11&	94.1&	93.55&	93.47	&94.54&94.52	\\
PS&\textbf{2.16}&	2.18&	2.36&	2.36&	2.26&	2.27&	2.29&	2.29&	2.29&	2.29&	2.25&2.25\\\hline
    \end{tabular}
    \label{tab:results}
\end{table}
The best averaged pinball score was obtained for the ERA method applied to asinh transformed prices. Also the averaged coverage probabilities are in this case closer to $5\%$ or $95\%$ for the expectile-based methods. 
We observe much improvement of the forecast accuracy if the asinh transformation is applied to electricity prices, especially in the lower tails of the predicted distribution and in the overall pinball score. Interestingly, if no transformation is applied, then the quantile-based methods yield higher accuracy than their expectile analogues. Those methods rely on the absolute deviation instead of the least squares, so are more robust to outliers. %Hence, if no variance stabilizing transformation is applied to electricity prices, then the expectile-based methods should be applied with consciousness.      
Nevertheless, accuracy of the forecasts obtained without transformation is lower then their transformed versions for all of the considered methods.

\section{Conclusions}
\label{sec:conclusions}
In this paper we proposed a new method for probabilistic forecasting of electricity prices. It is based on combining forecast averaging with the expectile regression. Precisely, it yields forecasts of expectiles, given as linear combinations of a pool of point forecasts. Predicted distribution is then given in terms of expectiles, so it can be directly used e.g. for risk management purposes. On the other hand, from a grid of expectiles one can also calculate quantiles of the same distribution. Such transformation yields prediction intervals, commonly used in the forecasting context. 

The proposed ERA approach was applied to the German electricity market data. Its accuracy for hourly, day-ahead electricity prices was compared with the QRA as well as the historical simulation methods. In terms of the pinball score both considered forecast averaging methods, ERA and QRA, significantly outperformed historical simulation. Results of the expectile- as well as quantile-based historical simulation methods were in this case similar. We also calculated coverage probabilities at the $5\%$ and $95\%$ levels. For this accuracy measure all expectile-based methods outperformed significantly the quantile-based approaches. Overall, the best results were obtained for the ERA method applied to prices after variance stabilizing transformation. Such transformation improved forecast accuracy for all considered methods.

We believe that utilizing the notion of expectiles in probabilistic forecasting of electricity prices might improve the forecasts accuracy. Since using expectile regression leads to least squares optimization, it naturally inherits its good numerical properties. However, for such volatile data as electricity prices it should be applied with consciousness, as the least squares method is not robust to outliers. Hence, a variance stabilizing transformation or outlier treatment methods might be necessary to apply it efficiently.      

\backmatter

\bmhead{Acknowledgements}
The research was financed by the Polish National Science Center (NCN) Sonata grant No. 2019/35/D/HS4/00369.

\section*{Statements and Declarations}
The authors declare no conflict of interest.
%Some journals require declarations to be submitted in a standardised format. Please check the Instructions for Authors of the journal to which you are submitting to see if you need to complete this section. If yes, your manuscript must contain the following sections under the heading `Declarations':
%\begin{itemize}
%\item Conflict of interest/Competing interests (check journal-specific guidelines for which heading to use)
%\item Ethics approval and consent to participate
%\item Consent for publication
%\item Data availability 
%\end{itemize}

%\noindent
%If any of the sections are not relevant to your manuscript, please include the heading and write `Not applicable' for that section. 

%%===================================================%%

\bibliography{bibliografia}

\end{document}